\documentstyle[preprint,aps,floats,epsfig]{revtex}
\begin{document}
\title{Enhanced magnetic fluctuations in doped spin-Peierls
systems: a single-chain model analysis} 
\author{Michele Fabrizio and R\'egis M\'elin }
\address{
International School for Advanced Studies (SISSA) Via Beirut 2--4,
34014 Trieste, Italy}
\maketitle
\begin{abstract}
We analyze by means of real space Renormalization Group (RG) as well as
by exact diagonalizations  
the properties of a single-chain model of a doped 
spin-Peierls system, where a major role is played by the localized 
moments created
by the impurities.
We are able to follow analytically the RG flow, which allows us to
determine the relevant cross-over temperatures. In particular, we
find an enhancement of magnetic correlations due to disorder, coexisting with
an underlying dimerization, in an 
intermediate temperature range below the spin-Peierls critical temperature
and above the coherence temperature of a regular array built by those localized
moments (so-called soliton bandwidth). The possible relevance of these 
results to the doped inorganic spin-Peierls compound CuGeO$_3$ is discussed. 
\end{abstract}

\section{Introduction}

The behavior of spin-Peierls systems in the presence of disorder
has recently attracted considerable interest in the light of the
intriguing properties of the inorganic CuGeO$_3$ compound.
The pure compound is a quasi-one dimensional material,
in which the interchain couplings in the two directions
perpendicular to the chains are estimated to be 10\% and 1\% 
of the intrachain coupling. 
At at temperature $T_{SP}\simeq 14K$, there is a structural
transition, below which the CuO$_2$ chains dimerize and a gap
in the spin excitation spectrum opens\cite{exp}. The properties below
$T_{SP}$ are reasonably well described by a dimerized Heisenberg chain,
with an additional next-nearest neighbour exchange coupling\cite{Emery}.

If few percent of Cu is substituted by magnetic (Ni\cite{Ni}) or 
non magnetic (Zn\cite{Zn-a,Zn-b,Zn-c}) ions, or, if Ge is replaced by
Si\cite{Si-a,Si-b}, besides the structural transition, which still
occurs close to 14K,
a second transition is detected at $T_N\simeq 2\div 4K$, which 
has been identified as a N\'eel antiferromagnetic transition.
The staggered magnetic moment with 4\% of Zn is estimated
of the order of 0.2$\mu_B$\cite{Zn-b}.

The simplest explanation of the appearence of a N\'eel phase
upon such a weak doping, is that already the pure compound is quite close
to a transition from a spin liquid phase,
with a gap in the excitation spectrum, to an antiferromagnetic   
state, with gapless spin wave excitations.
Then, one
could imagine that doping effectively reduces the dimerization, 
or increases the interchain coupling, so that the system is pushed
quickly in the magnetically ordered phase. Within this picture,
one would represent the disordered system still as a homogeneous
system but with modified parameters, and the enhancement
of the magnetic fluctuations would be simply due to the critical 
quantum fluctuations.
Notice that a dimerized exchange is not incompatible with 
a N\'eel order in more than one dimension, which explains the
coexistence of dimerization and magnetic ordering.
However, in our opinion, this explanation is not fully consistent
with the experiments.
A detailed discussion, including also a review of existing 
results, is postponed to the conclusion section. 

Therefore we believe that something else has to be invoked to
explain why the spin-liquid phase of the pure CuGeO$_3$ is
so unstable upon doping. 
In this paper we introduce and study another mechanism
of enhancement of the antiferromagnetic fluctuations, which
is purely due to disorder, exists also in a single chain, and can not
be represented by a homogeneous system with modified parameters.
Essentially, we will assume that, at low doping, the impurities
release spin-1/2 solitonic excitations, which, however, are not free
to move but get trapped in the vicinity of the impurities 
by interchain correlations\cite{Khomskii,nota-imp}.
Antiferromagnetic fluctuations can then be established since
a coupling between these localized spins
is generated by the polarization of the spin-singlet background.
However, one would naively disregard the effects of these
magnetic fluctuations, since they are supposed to appear at an
energy/temperature scale $T_*$ of the order of the
spin-wave bandwith which would be obtained if these
localized spins formed a regular lattice (soliton lattice bandwidth). 
This scale is proportional to the spin-Peierls gap with
a prefactor exponentially small in the ratio of the
average distance between the impurities to the soliton width.
At 2\% Zn-doping, with the estimated soliton width of 13.6 lattice
spacings\cite{soliton}, this prefactor would be of order 0.025,
while it would be 0.16 at 4\% doping. Since at 2\% Zn-doping
$T_N\simeq 0.28 T_{SP}$\cite{Zn-b}, it would be hard to believe
that these localized spins have something to do with the
antiferromagnetism. 
We will show that this naive conclusion is not correct when
these solitons are randomly distributed.
In fact, as a consequence of disorder, sensible
magnetic correlations will form well above $T_*$.
Therefore, this effect may in principle play a role
in the establishment of antiferromagnetism.  
Notice that this source of enhancement of magnetic fluctuations 
is not incompatible with the previously discussed scenario,
rather it favours the approach to the critical point separating
the spin liquid from the magnetically ordered phase.  
As we are going to discuss, we believe that there are evidences
that these disorder-induced fluctuations play a role in these materials, 
especially at low doping.

The paper is organized as follows. 
In section II we will introduce the model and discuss some
existing results. In section III we will introduce the
renormalization group approach which we use to study the model, and 
start the analysis, which will be further improved in section IV
with the calculation of the spin-spin correlation function.
In section V we present some numerical results which support
the renormalization group analysis. Section IV is devoted to
exact diagonalizations of small clusters,
the interchain coupling being treated in mean field.
The conclusions are given in section VII.

\section{The model}

The Hamiltonian which we consider in the absence of impurities
is simply that of a dimerized Heisenberg chain
\begin{equation}
\hat{H} = \sum_i (1 + \delta (-1)^i) \left( S^x_{i}S^x_{i+1} +
S^y_{i}S^y_{ i+1} + \Delta S^z_{ i}S^z_{ i+1} \right),
\label{H}
\end{equation}
where $\delta$ is the strength of the dimerization. As we said,
to correctly describe the CuGeO$_3$ compound,
one should include a next nearest
neigbour coupling. However, for our purposes, this is not really
important. 
We assume that one impurity releases one spin-1/2 solitonic excitation,
connecting regions of different dimerization parity\cite{notaSi}. The role
of the interchain coupling is to provide a confining potential to the 
soliton, which will be trapped within some distance from the 
impurity\cite{Khomskii} (as well as to enlarge the soliton 
width\cite{Bishop}). Moreover,
the weak link connecting the impurity nearest neighbors (which would be
generated by virtual hoppings into and out from the impurity site, as well as
by a next-nearest neighbor exchange) is  
approximated to be equal to the weak bonds in (\ref{H}). This approximation 
is valid if one is interested really in what happens close to
the middle of the spin-Peierls gap.
Therefore, for a finite number $n_{imp}$ of randomly distributed impurities, 
the effective model will be assumed to consist of a 
{\sl squeezed} chain with $n_{imp}$ sites less, described by the
same Hamiltonian Eq.(\ref{H}), but in the presence of 
randomly distributed domain walls. Their number will
be $\leq n_{imp}$, since we can not exclude that pairs of
solitons recombine. This amounts to take
a site dependent $\delta(i)$, which takes alternatively two values 
$\pm \delta$, jumping from one to the other at the (random) position
of the antiphase walls. If we assume no correlation between
the impurities, then the distance $r$ between two consecutive 
domain walls is distributed according to a Poisson law
\begin{equation}
P(r) = n {\rm e}^{-(r-a)n}\theta(r-a),
\label{Pr}
\end{equation}
where $n$ is the concentration of domain walls, and $a$ 
is of the order of the lattice spacing.
The XY version of this model [$\Delta=0$ in Eq.(\ref{H})],
has been recently studied in the continuum limit ($a\to 0$)
in Ref.\onlinecite{nous}. In this reference, exact expressions of
the density of states as well as of several thermodynamic quantities
have been derived. These results are supposed to give a good
qualitative description up to the isotropic point
[$0\leq \Delta\leq 1$ in Eq.(\ref{H})], since, as discussed in 
Refs.\onlinecite{Hirsh} and \onlinecite{Fisher}, the spin anisotropy 
in a disordered chain does not play the same crucial role as in the
absence of disorder. In this paper, we will study 
the low energy (lower that the spin-Peierls gap) 
behavior of the spin-isotropic model by
means of the real space Renormalization Group (RG) 
originally introduced by Dasgupta and Ma\cite{Ma} to cope with 
random spin chains, and further extended
by Fisher\cite{Fisher}. We will show that this approach reproduces
the exact results obtained in the XY limit in Ref.\onlinecite{nous},
thus proving the unimportance of the spin anisotropy
for $0\leq \Delta\leq 1$.
In addition, by this technique it is also possible to calculate
the spin-spin correlation functions, which were not calculated 
in Ref. \onlinecite{nous}, being not easily
accessible by the method used in that paper.

\subsection{Effective low energy model}

We start our analysis by building an effective low energy model
which should describe the Hamiltonian (\ref{H}) below the
spin Peierls gap, and for low doping. In this regime, most of the spins are
frozen into singlets apart from the 
spin-1/2 solitons localized around the domain walls. 
Two consecutive solitons are coupled by an exchange which we assume of the
form
\begin{equation}
J(r) = \phi_0 {\rm e}^{-r\phi_0},
\label{Jr}
\end{equation}
where $r$ is the distance between the two, which is distributed according to
(\ref{Pr}), and $1/\phi_0$ is the soliton width. We have choosen units in 
which the energy has the dimension of an inverse length. 
Therefore the resulting low energy model describes a random Heisenberg model 
where spins are randomly 
distributed on a chain according to the law (\ref{Pr}), and coupled by 
the exchange constants (\ref{Jr}). This model should give
a correct description of (\ref{H}) at energies/temperatures smaller than
$\phi_0$, and for $n\ll \phi_0$ (the solitons should not overlap too much),
and can be analysed by the RG technique of Ref.\onlinecite{Fisher},
as we are going to show in the following sections.
A similar low energy picture has been proposed in Ref.\onlinecite{Dagotto}
on the basis of an exact diagonalization of a realistic model for CuGeO$_3$,
although limited to small clusters. By analyzing their data,
the authors of that reference have concluded that the random Heisenberg model
is indeed a good description for the low energy excitations. 
Here, given for granted this low energy picture, we analyze 
some possible consequences, which can be useful 
for a comparison with experimental data as well as a guiding line 
for further numerical studies.

It is now worthwhile to realize that the antiferromagnetic correlations
in the squeezed chain do not simply translate into oscillations with
wave vector $Q=\pi/a$ in the original chain. This is due to the
presence of the impurities which change one sublattice into the other,
since the coupling across the impurity site is antiferromagnetic.
For instance, if in the squeezed chain a correlation function oscillates
like $\cos(Qx)$, in the original chain it will oscillate
like $\cos[Q(x-N(x))]$, where $N(x)$ is the number of impurities
in that interval. For randomly distributed impurities, after the
averaging, this behavior will transform into
$\cos(Qx){\rm e}^{-2n_i x}$, where $n_i$ is the impurity concentration.
Therefore, within this single chain model, we can not describe a true N\'eel 
long range order. For that, we would need to properly take into account the
interchain coupling. A first attempt is discussed in section IV.C, where
we calculate the susceptibility to a staggered (in the original chain, not
in the {\it squeezed} one) magnetic field. 
However, for the purpose of describing how disorder enhances magnetic
fluctuations, without pretending to push the analysis up to the 
AF transition, the single chain model we study is sufficient.

To conclude, we notice that the XY version of this model is equivalent
to a tight-binding Hamiltonian for spinless fermions at half filling
with random hopping integrals.
This model has been studied quite a lot in connection with the
Anderson localization, for its intriguing properties.  
A detailed analysis was carried out by Eggarter and 
Riedinger\cite{Eggarter}. They showed, quite generally, that
the density of states has a Dyson singular behavior at the
chemical potential 
$\rho(\epsilon) \sim 1/|\epsilon\ln^3\epsilon|$, and that the localization 
length diverges logarithmically $\lambda(\epsilon)\sim |\ln\epsilon|$.
However, they did not discuss the consequences of these singularities
in the correlation functions. This analysis was partially 
carried out by Gogolin and Mel'nikov\cite{Gogolin}
using the Berezinskii diagram technique. They showed, for instance,
that, although one dimensional, this system has a finite conductivity.
However, their approach is quite complicated and does not provide
a simple interpretation of the low energy behavior. 
On the contrary, the RG analysis carried out by Fisher for the
spin model is, physically, more simple, and, once translated in the fermion
language of the tight-binding model, provides not only a
simple low energy picture but also many new results, as 
for instance the zero temperature
power law decay of the average density-density
correlation function. 
We have verified these predictions numerically in the tight-binding
model, as we are going to discuss in section V.

\section{Renormalization of the bond distribution}
\label{bonddis}

We start giving a very short introduction to the RG transformation. 
Then, as a warm up exercise, we 
will calculate the renormalization of the probability distribution
of the bond exchange-couplings. 

The RG transformation consists in successive eliminations
of the most strongly coupled pairs of spins, by projecting out the
Hilbert space onto the subspace where those pairs are frozen into 
singlets. The cut-off energy which is rescaled downwards in the course of
the renormalization procedure is the strength of the strongest bond
$\Omega=\max\{J\}$. Therefore, going from $\Omega$ to 
$\Omega - \delta \Omega$, amounts to project out 
all those pairs of spins which are coupled by an exchange of value
$\Omega$. This operation generates an effective
bond between two spins separated by a projected pair.
If, for instance, at the energy scale $\Omega$, the
coupling between spins 2 and 3 is $J_{2,3}=\Omega$,
then the projection onto the subspace in which
those spins are frozen into a singlet generates a
coupling between the neigbouring spins 1 and 4, given by
$J_{1,4} = J_{1,2} J_{3,4}/J_{2,3}$. The decimation scheme
becomes more and more valid in the course of the scaling procedure,
as one verifies by the flowing of the bond probability distribution.

In our particular case, as previously discussed, we will consider
a chain where spins are randomly distributed in such a way that the
distribution of distances $r$ between two consecutive spins
is given by (\ref{Pr}). The bond connecting two consecutive spins 
at distance $r$ has a strength given by (\ref{Jr}).
Following Fisher, we define $\Gamma= \ln(\phi_0/\max\{J\}) $, and
a new variable $\zeta = - \Gamma + \ln(\phi_0/J(r))$.
The recursion relation for the $\zeta$-variable is quite simple
in the above decimation scheme. 
In fact, in our previous example,
where the sites 2 and 3 are decimated,
a coupling between sites 1 and 4 is generated, such that
 $\zeta_{1,4} = \zeta_{1,2} + \zeta_{3,4} - \zeta_{2,3} 
= \zeta_{1,2}+\zeta_{3,4}$, being $\zeta_{2,3}=0$. We also 
define a scaling variable
$\eta=\zeta/\Gamma$.
At our RG starting point, $\Gamma_0=\phi_0 a$,
$\eta = (r-a)/a$, 
and the initial $\eta$ distribution is
\[
Q(\eta,\Gamma_0) = \frac{\Gamma_0 n}{\phi_0}
\exp{\left(- \frac{n \Gamma_0}{\phi_0} \eta \right)}
\theta(\eta).
\]
With these notations, the RG scaling equation for $Q$ is\cite{Fisher}
\begin{equation}
\Gamma \frac{\partial Q}{\partial \Gamma}
= Q + (1+\eta) \frac{\partial Q}{\partial \eta}
+ Q(0,\Gamma) \int_0^{\eta}
d \eta' Q(\eta',\Gamma) Q(\eta-\eta',\Gamma).
\label{RG-eq}
\end{equation}
We search for a solution of (\ref{RG-eq}) of the form:
$Q(\eta,\Gamma) = \bar{f}(\Gamma) \exp{(- 
\bar{f}(\Gamma) \eta)}
\theta(\eta)$, with $\bar{f}$ independent of $\eta$.
At the starting point $\bar{f}(\Gamma_0)  = n a$.
The equation of motion of $\bar{f}(\Gamma)$ is 
\begin{equation}
- \Gamma \frac{\partial
\bar{f}(\Gamma)}{\partial \Gamma}
= - \bar{f}(\Gamma) +
\bar{f}^{2}(\Gamma),
\end{equation}
whose solution compatible with the boundary condition is
\begin{equation}
\label{barf}
\bar{f}(\Gamma) = \frac{\Gamma n}
{\phi_0(1-n a) + \Gamma n}.
\end{equation}
If $\Gamma=a\phi_0$, we recover the initial $Q(\eta,\Gamma_0)$
distribution, and, in the fixed point limit, $\Gamma \rightarrow
+ \infty$, we have $Q^{*}(\eta)=\exp(-\eta)\theta(\eta)$, which is exactly
the random-singlet fixed point analyzed in Ref.\onlinecite{Fisher}.
In addition, we have an analytic expression of $Q$ at all intermediate
scales. 

If $n(\Gamma)$ denotes the number of spins per unit length
not yet decimated at scale $\Gamma$,
we have \cite{Fisher}
\begin{equation}
\frac{d n(\Gamma)}{n(\Gamma)}
= - \frac{2 Q(0,\Gamma)}{\Gamma} d \Gamma
.
\end{equation}
Using the previous solution for the bond distribution,
taking the limit $a \rightarrow 0$, and denoting
$\Gamma = \ln{(\phi_0/E)}$ ($E\leq \phi_0$), we have
\begin{equation}
n(E) = n \left( \frac{\phi_0}
{ n \ln{(\phi_0/E)} + \phi_0}
\right)^{2}.
\end{equation}
At the begining of the renormalization, $E=\phi_0$,
and we have $n(\phi_0)=n$, as expected. In the small
$E$ limit
\begin{equation}
n(E) \simeq \frac{\phi_0^{2}}
{n\ln^2(\phi_0/E)}.
\end{equation}
This leads to a low temperature uniform susceptibility
\begin{equation}
\chi(T) = \frac{n(T)}{4 T} = \frac{\phi_0^{2}}
{4n T \ln^2{(\phi_0/T)} },
\label{unif-susce}
\end{equation}
which is exactly the low temperature
susceptibility obtained in Ref. \onlinecite{nous} for the XY limit
of a dimerized chain with a dilute random distribution of domain walls.
The agreement between the two results is, first of all, a
proof that the two models are equivalent at low energies (smaller than
the spin-Peierls gap $\phi_0$), and, secondly, an evidence that
the spin-anisotropy is not relevant.

\section{Bond length distribution}
\label{length}
We now analyze the joint probability distribution of
bond lengths and couplings, which allows to calculate
the spin--spin correlations at arbitrary energy scale, and thus
the relevant cross--over temperatures.

Let $P(\zeta,l,\Gamma)$ be the joint probability distribution
of a bond of length $l$ and coupling $\zeta$ at scale $\Gamma$. Following
Fisher \cite{Fisher}, the RG transformation of
this quantity is
\begin{eqnarray}
& &\frac{\partial P(\zeta,l,\Gamma)}{\partial \Gamma}
= \frac{\partial}{\partial \zeta} P(\zeta,l,\Gamma)\label{RG-eqbis}\\
& &+ \int_0^{+ \infty} d l_1 d l_2 d l_3 
d \zeta_1 d \zeta_3 \delta(\zeta - \zeta_1 - \zeta_3)
\delta(l-l_1-l_2-l_3) P(0,l_2,\Gamma)
P(\zeta_1,l_1,\Gamma) P(\zeta_3,l_3,\Gamma)
\nonumber.
\end{eqnarray}
For the Laplace transform of $P(\zeta,l,\Gamma)$,
\begin{equation}
\hat{P}(\zeta,y,\Gamma) = \int_0^{+ \infty} dl e^{-ly}
P(\zeta,l,\Gamma),
\end{equation}
we look for a solution
\begin{equation}
\label{ansatz}
\hat{P}(\zeta,y,\Gamma) = \Phi(y,\Gamma) {\rm e}^{- 
f(y,\Gamma) \zeta} \theta(\zeta)
\end{equation}
with $\Phi$ and $f$ two $\zeta$-independent functions.
By (\ref{RG-eqbis}), these two functions are solutions of the 
following differential equations
\begin{eqnarray}
\label{eq1}
\frac{\partial f(y,\Gamma)}{\partial \Gamma}
&=& - \Phi^{2}(y,\Gamma),\\
\frac{\partial \Phi(y,\Gamma)}{\partial \Gamma}
&=& - f(y,\Gamma) \Phi(y,\Gamma).
\label{eq2}
\end{eqnarray}
Notice that (\ref{ansatz}) is for the moment just an ansatz,
which is valid only if one is able to solve 
(\ref{eq1}) and (\ref{eq2}) consistently with the appropriate boundary 
conditions. 
For our particular model, the initial joint probability distribution is 
\begin{equation}
P(\zeta,l,\Gamma_0) = n e^{-n(l-a)} \delta\left(\zeta-(l-a)\phi_0\right)
\theta(\zeta).
\end{equation}
After Laplace transforming with respect to $l$, we
find the following boundary conditions ($a\to 0$) 
\begin{eqnarray}
f(y,\Gamma_0) &\equiv& f_0(y) = \frac{y+n}{\phi_0}, \label{bc-a}\\
\Phi(y,\Gamma_0) &=&  \frac{n}{\phi_0}. \label{bc-c} 
\end{eqnarray}
Solving Eqs.(\ref{eq1}) and (\ref{eq2}), we obtain 
\begin{eqnarray}
f(y,\Gamma) &=& \sqrt{C(y)}\, \frac{ f_0(y) \cosh(\sqrt{C(y)}\Gamma)
+ \sqrt{C(y)}\sinh(\sqrt{C(y)}\Gamma) }{ f_0(y) \sinh(\sqrt{C(y)}\Gamma)
+ \sqrt{C(y)}\cosh(\sqrt{C(y)}\Gamma)},\label{f}\\
\Phi(y,\Gamma) &=& \frac{n}{\phi_0}  \sqrt{C(y)}\, 
\frac{1}{ f_0(y) \sinh(\sqrt{C(y)}\Gamma)
+ \sqrt{C(y)}\cosh(\sqrt{C(y)}\Gamma)}\label{Phi},
\end{eqnarray} 
where $C(y)=f^{2}(y,\Gamma_0) - \Phi^2(y,\Gamma_0)
 = y(y+2n)/\phi_0^2$. 
One can verify that the ansatz solution (\ref{ansatz}) with the 
two functions defined by (\ref{f})-(\ref{Phi}) does satisfy the appropriate
boundary conditions. Moreover, when $y=0$, we do recover the 
$\zeta$-probability distribution found in the previous section. If we scale
$y=\tilde{y}\phi_0^2/(2 n\Gamma^2)$, and send $\Gamma\to\infty$ keeping
$\tilde{y}$ constant, we find
\begin{equation}
\hat{P}\left( \frac{\tilde{y}\phi_0^2}{2n\Gamma^2},\Gamma\eta,\Gamma\right)
\to \frac{\sqrt{\tilde{y}}}{\Gamma\sinh\sqrt{\tilde{y}}}
{\rm e}^{-\eta\sqrt{\tilde{y}}\coth\sqrt{\tilde{y}}},
\label{fixed-point}
\end{equation}
which is the fixed point distribution found in Ref.\onlinecite{Fisher}. 
The joint probability distribution $P(l,\zeta,\Gamma)$ is obtained
after the inverse Laplace transformation:
\begin{equation}
P(l,\zeta,\Gamma) = \int_{c-i\infty}^{c+ i \infty}
\frac{dy}{2\pi i} {\rm e}^{ly} \Phi(y,\Gamma)
{\rm e}^{-f(y,\Gamma)\zeta}\theta(\zeta).
\label{sol-partial}
\end{equation}
Having calculated $P(l,\zeta,\Gamma)$, we are now in position to derive 
various important quantities, as we are going to show. 

\subsection{Fixed-point bond length probability distribution}
The bond length probability distribution $P(l,\Gamma)$ 
at the fixed point is obtained
by integrating (\ref{sol-partial}) over $\zeta$, and 
assuming $y\ll n$. The result is\cite{Fisher}
\begin{equation}
P(l,\Gamma) = \frac{2\pi\phi_0^2}{n\Gamma^2} 
\sum_{m=0}^\infty \left(n+\frac{1}{2}\right) (-1)^n
{\rm e}^{-\pi^2(n+1/2)^2\tilde{l}/\Gamma^2},
\end{equation} 
where $\tilde{l}=l\phi_0^2/2n$ is the length in appropriate dimensionless
units.
This expression coincides with the probability distribution 
that a random walker remains inside an interval $\Gamma$ after
a ``time'' $\tilde{l}$. The same probability was found to play a
crucial role in the study of the one-dimensional 
tight-binding Hamiltonian with off-diagonal disorder\cite{Eggarter}. 
The comparison between the RG
approach and the analysis of the Schr{\oe}dinger equation
for the tight-binding Hamiltonian provides the quantum-mechanical 
interpretation of the bond length probability distribution 
at scale $\Gamma$ as the probability distribution of half of the distance
between two consecutive nodes of the wavefunction at energy 
$E=\phi_0\exp{(-\Gamma)}$. The average bond length or, equivalently, half
the average distance between the nodes at energy $E$, is simply given by
\[
<l> = \frac{2 n \ln^2(\phi_0/E)}{\phi_0^2},
\] 
from which one obtains an estimate of the integrated density of state
$N(E)=1/(2<l>)$, in perfect agreement with the exact result obtained for
the dimerized XY chain with randomly distributed domain walls\cite{nous}.

\subsection{Spin-spin correlation functions}

The spin-spin correlation function 
\[
\chi_a(l) = \langle S^a(0) S^a(l) \rangle,
\]
($a=x,y,z$), can be also calculated by making
use of the joint probability distribution. At finite temperature
$T\leq \phi_0$, the RG has to be stopped at an energy scale 
$\Gamma_T = \ln(\phi_0/T)$. If a bond of lentgh $l$ has a 
coupling larger than $T$, it will be decimated, and 
contribute a constant to $\chi(l)$. On the contrary, for the
bonds of the same length but coupling smaller than $T$, 
one can, as a first approximation, perform a high temperature expansion,
keeping only the first non vanishing term. 
As a result, we assume the following expression
for the spin-spin correlation function:
\begin{eqnarray}
\chi(l,T) &\simeq& \int d\zeta \frac{J(\zeta)}{T}
n(\Gamma_T) P(\zeta,l,\Gamma_T) + 
\int_{\Gamma_0}^{\Gamma_T} d\Gamma n(\Gamma) P(0,l,\Gamma)
\nonumber \\
&=& n(\Gamma_T)\int d\zeta {\rm e}^{-\zeta} P(\zeta,l,\Gamma_T)
+ \int_{\Gamma_0}^{\Gamma_T} d\Gamma n(\Gamma) P(0,l,\Gamma)
\equiv \chi_1(l,T) + \chi_2(l,T),
\label{chi-def}
\end{eqnarray}
where $\chi_1$  is the result of the high temperature expansion
(notice that, by definition, $\zeta=\ln(T/J)$ at scale $\Gamma_T$), 
and $\chi_2$ is the contribution of the decimated bonds. 
Notice that, by the way in which the decimation scheme is built, 
the distance of a bond connecting two spins is always odd in
units of the lattice spacing, at least in the {\it squeezed} chain.
Therefore the above correlation function is in fact staggered
in the {\it squeezed} chain. 
 
Since a very long bond will most likely form at scale 
$\Gamma<\Gamma_T$\cite{Fisher}, $\chi_1$
will dominate the correlation function when $l$ is very large
and for intermediate $\Gamma_T$.
(On the contrary for $\Gamma_T\to\infty$ it is 
$\chi_2$ which dominates.) 
In this limit, $\chi\simeq \chi_1$
slightly differs from the expression used by Fisher,
i.e. $\chi(l)\sim P(0,l,\Gamma_T)$.
In fact, the difference is irrelevant at low
temperature, but is important at higher temperature.
In particular, with the definition (\ref{chi-def}), 
we can recover at $T\sim \phi_0$ the correct asymptotic behavior 
$\chi(l)\sim \exp{(-\phi_0 l)}$, for $\phi_0\gg n$. Since we are 
mostly interested in the cross-over, we will use (\ref{chi-def}).

\protect\subsubsection{Thermal correlation length}
Let us first discuss the behavior of $\chi_1$ in (\ref{chi-def}).
After the $\zeta$-integration, one has to calculate
\begin{equation}
\chi_1(l,T) = n(\Gamma_T)  
\int_{c-i\infty}^{c+i\infty} \frac{dy}{2\pi i}
\frac{\Phi(y,\Gamma_T) {\rm e}^{ly}}{
1 + f(y,\Gamma_T)}.
\label{chi1}
\end{equation}
At finite temperature, the long distance behavior will be dominated
by the nearest singularity to the origin in the complex $y$-plane.
This pole will define the inverse of the thermal correlation length
$\xi_T$, by $\chi(l)\sim {\rm e}^{-l/\xi_T}$.
We will compare the resulting $\xi_T$ with the thermal correlation length
$\xi_{0T}$ of a chain where spins are regularly distributed with interspin  
distance $1/n$ and coupled by an
exchange $J_0 =\phi_0 {\rm e}^{-\phi_0/n}$ (soliton lattice). 
For convenience, we
will consider the longitudinal spin-spin correlation function in
the XY limit of both models. This amounts to 
substitute 1 with 2 in the denominator of Eq.(\ref{chi1}).
This comparison is useful to understand in which regime the 
spin-spin correlation is reduced by disorder, and in which regime, if any,
is on the contrary enhanced.
We find that
\begin{equation}
\xi_{0T} = \frac{1}{2n\ln\left( \frac{\pi T}{2 J_0}
+ \sqrt{\left(\frac{\pi T}{2 J_0}\right)^2 + 1} \right)}.
\label{xi-free}
\end{equation}
If $T\ll T_*$, where $T_* = 2J_0$ is the
bandwidth of the spin excitations, then
\begin{equation}
\xi_T = \frac{2n\ln^2(\phi_0/T)}{\pi^2\phi_0^2}.
\label{xi-smallT}
\end{equation}   
The correlation length diverges as $T\to 0$, even though
more slowly than for the regular lattice, where 
$\xi_{0T}\sim J_0/(\pi nT)$. Therefore, for very low temperature, we
find that the disordered system has weaker correlations than the
ordered soliton lattice, a result rather obvious.    

\begin{figure}
\epsfig{file=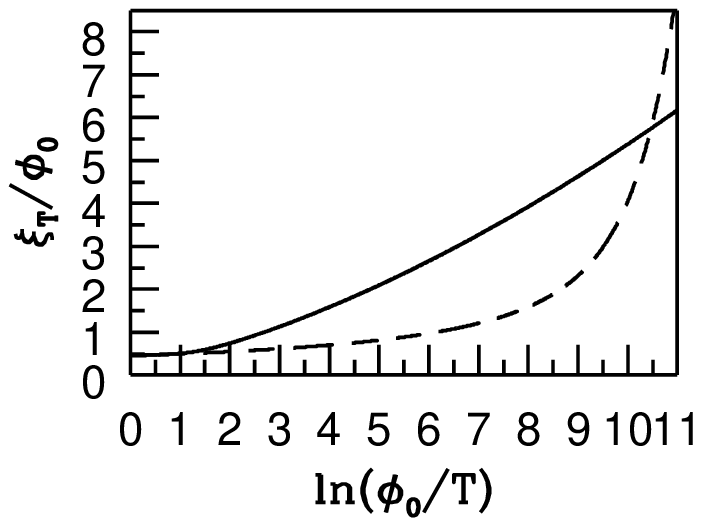}
\vspace{-3cm}
\caption{Thermal correlation lengths of the disordered system 
with $n/\phi_0 = 0.1$ (solid line), and 
for the non disordered one (dashed line).}
\label{Fig1}
\end{figure}

On the contrary, within the interval $T_*<T<\phi_0$, a different situation 
is encountered, in which the correlation length of the disordered 
system is larger that the one of the ordered lattice, as shown in Fig.
\ref{Fig1}. We find that the correlation
length of the disordered system increases linearly in $\ln(\phi_0/T)$
in that temperature range, while $\xi_0$ remains almost constant.
This result is also somewhat predictable, even though at first glance 
surprising. In fact, above the
coherence temperature $T_*$, the disordered system is
more correlated since it takes advantage from 
configurations where the spins are closer than in an ordered lattice.
In connection with our original model of a doped spin-Peierls
system, this result implies that magnetic correlations may appear
at temperatures well above the soliton bandwidth, which is not
in contradiction with the experimental evidences in CuGeO$_3$.

Let us now consider the limit $T\to 0$ , where the contribution of
$\chi_2$ in (\ref{chi-def}) is dominant. With exponential accuracy, we can
use the fixed-point joint probability distribution (\ref{fixed-point}).
After inverse Laplace transform,
\[
P(0,l,\Gamma) = \frac{\pi^2\phi_0}{2\Gamma^3}
\sum_{m=0}^\infty (-1)^{m+1} m^2 {\rm e}^{-\pi^2 m^2 \tilde{l}/\Gamma^2} .
\]
Performing the integral over $\Gamma$ and then summing the series, 
we find $\chi(l) \simeq n/(3\phi_0 l^2)$, which is the known power law
behavior\cite{Fisher} for a random Heisenberg chain. 

\subsection{Staggered magnetic susceptibility}

Another relevant quantity which can be calculated is the
staggered magnetic susceptibility. We add to the Hamiltonian 
a term $h\sum_R (-1)^R S_{zR}$, where $R$ is the localized spin
position in the original chain, and calculate the second order
correction to the free energy
\begin{equation}
\delta F(T) = -\frac{Th^2}{L} \int_0^\beta d\tau_1 \int_0^{\tau_1} d\tau_2
\sum_{R.R'} (-1)^{R+R'} << S_{zR}(\tau_1)S_{zR'}(\tau_2)>>,
\label{dF}
\end{equation}
where $<<\cdots>>$ denotes a thermal and disorder average, and
$L$ is the size of the chain.  
At scale $T$, mostly the spins which are still free will contribute 
to (\ref{dF}). Therefore, one
can again perform a high temperature expansion for these spins
and keep only the zeroth and first order term. 

At zeroth order, only $R=R'$ contributes to (\ref{dF}),
and gives
\[
\delta F^{(0)}(T) = -\frac{h^2}{8T} n(T),
\]
which results into a staggered susceptibility equal to the uniform
susceptibility of Eq.(\ref{unif-susce}). This clearly reflects the fact that
only on-site correlations are involved, which do not distinguish 
between uniform or staggered fields.

Longer range correlations start to appear at first order. We find   
\[
\delta F^{(1)}(T) = \frac{h^2}{16 T L}
\langle \sum_{R,R'} (-1)^{R+R'} \left( \frac{J_{R,R'}(T)}{T} \right)
\rangle ,
\]
where the remaining average is over the disorder, and $J_{R,R'}(T)$
is the exchange between the spins at sites  $R$ and $R'$ 
in the original chain at scale $T$, on
provision that these two spins have not been decimated yet.
At this point we have to be very careful in distinguishing the
original from the {\it squeezed} chain.
Notice that the number of spins which have been already decimated
and which lye in between the two spins is by construction even. If
we assume that each spin is localized very close to each impurity,
then also the number of impurities between $R$ and $R'$ will be
even. Therefore the parity of the distance $R-R'$ between the two spins 
depends whether they are at the right or the left of the impurities to 
which they are bound. In particular, 
the case in which the distance between the 
two spins in the squeezed chain is $l$ can correspond to 
four possible cases in the original chain. The first two cases, 
which correspond to $R-R'$ even, are
when $R$ and $R'$ are both on the right or on the left of the impurities to 
which they are bound.
We define these two cases, in obvious notations, as $(+,+)$ and $(-,-)$.
The other two cases, corresponding to $R-R'$ odd, occur when one spin is 
on the left and the other on the right and viceversa,
which we define as $(-,+)$ and $(+,-)$. 
The value of the coupling will
be different in the four cases. What enters in the calculation
of the staggered susceptibility is the combination
$J_{-,+} + J_{+,-} - J_{-,-} - J_{+,+}$ of the exchange
constants in the various cases. Due to the exponential dependence
of $J$ upon the distance, this combination will be 
always positive, as if $R-R'$ were effectively odd.
We therefore assume that
\[
(-1)^{R+R'} J_{R,R'}(T) = \gamma J_l(T) > 0,
\] 
where $J_l(T)$ is the exchange coupling of a bond of length $l$ 
in the {\it squeezed} chain at scale $T$, and $0<\gamma <1$ is a
reduction factor.
With this assumption we get
\begin{equation}
\delta F^{(1)}(T) =  \gamma \frac{h^2}{16 T }n(T)
\int dl d\zeta {\rm e}^{-\zeta} P(\zeta,l,\Gamma_T) =
\gamma \frac{h^2n(T)}{16 T} 
\frac{\bar{f}(\Gamma_T)}{\Gamma_T + \bar{f}(\Gamma_T)},
\label{dF1}
\end{equation}
where $\bar{f}$ has been defined in Eq.(\ref{barf}). At low temperature,
this gives a contribution to the staggered susceptibility 
$\sim 1/(T|\ln T|^3)$, which is subleading with respect to the
zeroth order term, but still diverging.

\section{Numerical calculations at zero temperature}
We now turn to the numerical calculations of the zero temperature
correlation length of our effective model. In the framework
of the renormalization group approach to the bond disordered model,
Fisher \cite{Fisher} showed that the zero temperature correlations
$<< S_i^{\alpha} S_j^{\alpha} >>$
behave like $1/|i-j|^{2}$, whatever $\alpha=x,y,z$, as we have seen
in the previous section. Our goal
is to check this prediction by means of exact diagonalizations
in the XY case.

We use the Jordan-Wigner transformation \cite{Lieb}
to map the spin $1/2$ 
problem to a spinless fermion problem. We work with an even number
of sites so that the fermions are periodic, and in the
$S_z=0$ sector, implying half-filling.
The XY Hamiltonian in the fermion language is
the one of non interacting fermions with random hoppings
\begin{equation}
H = \frac{1}{2} \sum_i J_i \left( c_{i+1}^{+} c_i
+ c_i^{+} c_{i+1} \right)
,
\end{equation}
where the $J_i$ exchanges are all antiferromagnetic
and distributed at random. We choose the bond distribution
$P(J) = \Theta (J) \Theta (1-J)$. For this model, we will
calculate both the $<< S_i^{z} S_j^{z}>>$
and $<< S_i^{+} S_j^{-}>>$ correlations.
This model belongs to the
same universality class as the model defined
by (\ref{Pr}) and (\ref{Jr}). In order to
test the equivalence between these two models,
we will calculate the $\langle \langle S_0^{z} S_R^{z}
\rangle \rangle$ correlations of the model (\ref{Pr})
and (\ref{Jr}) and show that it indeed decays like
$1/R^{2}$.

\subsection{$S^{z} S^{z}$ correlations}
The first step is to calculate numerically the spectrum
and the eigenvalues of the tight-binding Hamiltonian
\begin{equation}
\label{TB}
H = \frac{1}{2} \sum_i J_i \left(
| i+1 \rangle \langle i |
+ | i \rangle \langle i+1| \right)
.
\end{equation}
Let 
$
| \Psi^{\alpha} \rangle = \sum_i \Psi_i^{\alpha}
|i \rangle
$
the eigenstates of (\ref{TB}). The quantum average over the
half-filled Fermi sea (FS) is calculated as
$
<S_i^{z} S_j^{z}> = A_{i,j} - B_i / 2 - B_j / 2
$, with
\begin{eqnarray}
A_{i,j} &=& \sum_{\alpha,\beta \in FS}
\left( \Psi_i^{\alpha} \Psi_j^{\beta} \right)^{2}
+ \sum_{\alpha \in FS} \sum_{\beta \not{\in} FS}
\Psi_i^{\alpha} \Psi_i^{\beta} \Psi_j^{\beta}
\Psi_j^{\alpha}\\
B_i &=& \sum_{\alpha \in FS} \left( \Psi_i^{\alpha} \right)^{2}
.
\end{eqnarray}

\begin{figure}
\vspace{10cm}
\centerline{\epsfig{file=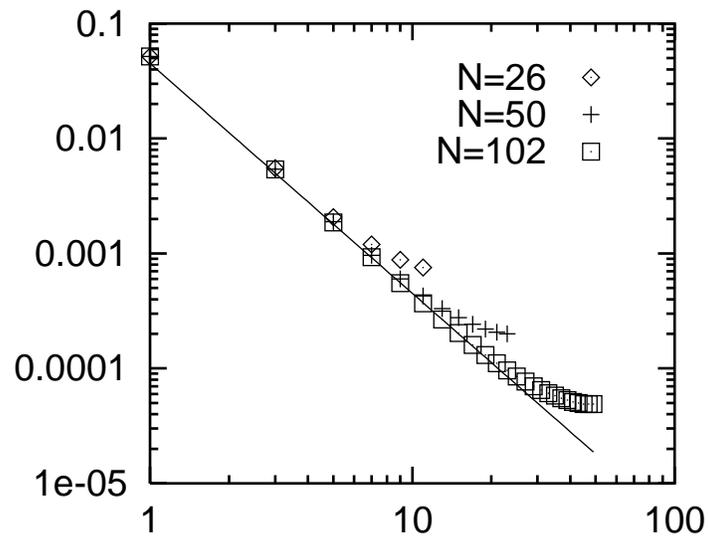,rwidth=10cm,rheight=10cm}}
\vspace{-5cm}
\caption{Average correlation for $N=26$, $N=50$ and $N=102$ sites.
200000, 350000 and 120000 disorder realizations were used
respectively. The solid line is the fit to the $1/x^{2}$
behavior.}
\label{Fig2}
\end{figure}

The results for the $<<S_i^{z} S_j^{z}>>$ correlation
are shown on Fig. \ref{Fig2}, in good agreement with
the $1/|i-j|^{2}$ scaling.

\begin{figure}
\vspace{10cm}
\centerline{\epsfig{file=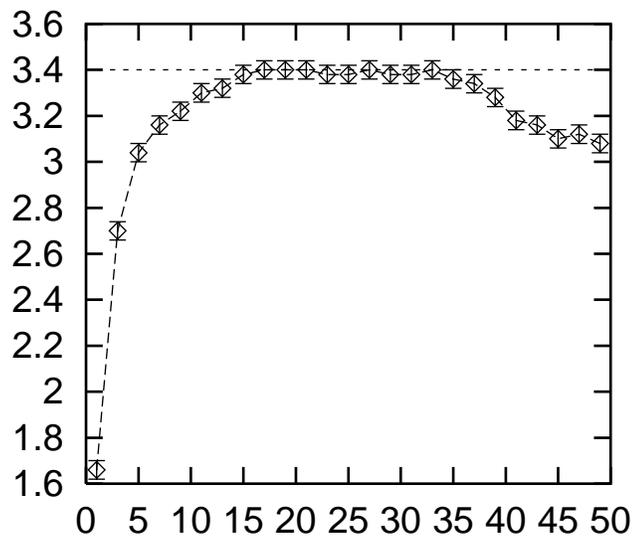,rwidth=10cm,rheight=10cm}}
\vspace{-5cm}
\caption{Behavior of the typical correlation as a function
of the separation. We have plotted $-\ln{|C^{typ}_{ij}|}
/|i-j|^{1/2}$ as a function of the spin separation, calculated
for a $N=102$ chain and 120000 disorder realizations.
A plateau is visible for distances between 20 and 35,
consistent with Fisher's prediction for the large
scale behavior of the typical correlations.
}
\label{Fig3}
\end{figure}

A main feature of the random singlet fixed point
is that typical and average correlations differ.
Fisher \cite{Fisher} claims that the typical
correlation behaves like $-\ln{|C^{typ}_{ij}|}
\sim |i-j|^{1/2}$. In order to check this prediction
numerically,
we calculated the typical $S^z S^z$ correlation function,
(defined as the
most probable correlation) as a function of distance.
The quantity $-\ln{|C^{typ}_{ij}|}/|i-j|^{1/2}$ is plotted
on figure \ref{Fig3}, where a plateau is clearly visible,
indicating a consistency between Fisher's prediction
and our numerical calculations.

\begin{figure}
\vspace{10cm}
\centerline{\epsfig{file=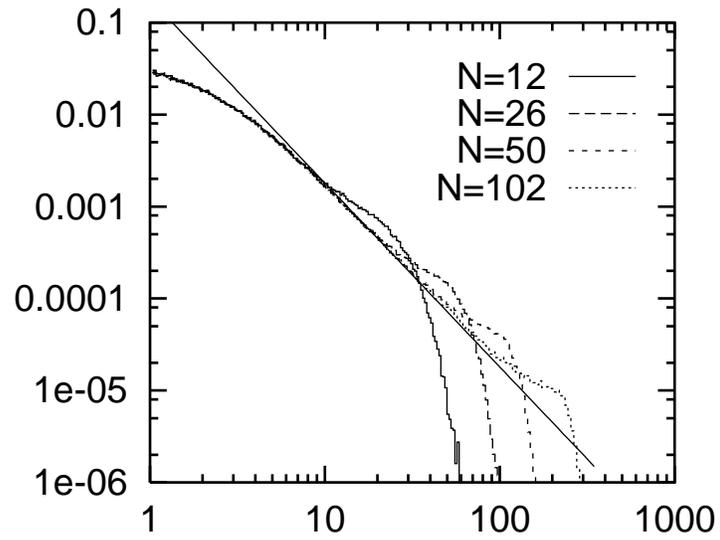,rwidth=10cm,rheight=10cm}}
\vspace{-5cm}
\caption{$S^{z} S^{z}$ correlations of the model
(\protect \ref{Pr}) and (\protect \ref{Jr}) for
$\phi_0=1$ and $1/n=5$, and
$N=12$,$N=26$, $N=50$ and  $N=102$ sites. 50000
disorder realizations were used excepted for $N=102$
where 13000 disorder realizations were used.}
\label{Fig4}
\end{figure}

In the case of the
(\ref{Pr}) and (\ref{Jr}) model, the $<<S_X^{z} S_{X+R}^{z}>>$
correlations
are shown on figure \ref{Fig4}, where a $1/R^{2}$ behavior
is also visible at large distances.

\subsection{$S^{+} S^{-}$ correlations}
\begin{figure}
\vspace{10cm}
\centerline{\epsfig{file=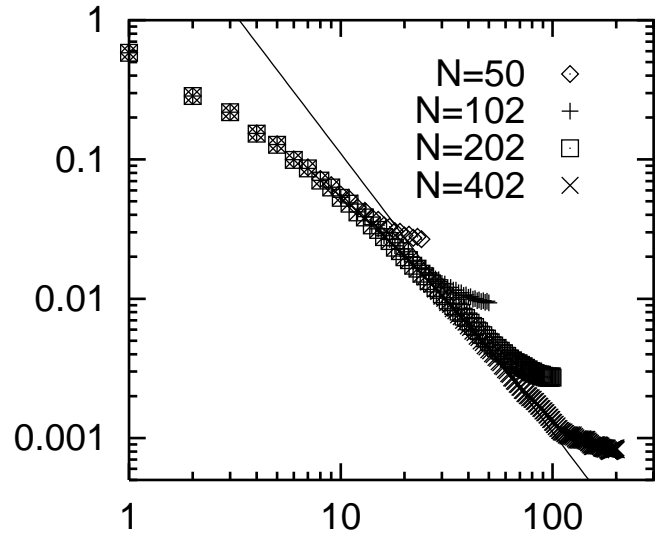,rwidth=10cm,rheight=10cm}}
\vspace{-5cm}
\caption{$S^{+} S^{-}$ correlations of the
disordered chain for $N=50$, $N=102$, $N=202$, $N=402$ chain.
The number of disorder realizations is
1000 except for the $N=402$ case where $420$
disorder realizations have been used. The solid line
indicates the $1/R^{2}$ behavior.}
\label{Fig5}
\end{figure}

We now turn to the computation of the $S^{+} S^{-}$
correlations of the XY chain. 
Following \cite{Young}, we define
$\hat{A}_i = c_i^{+} + c_i$ and 
$\hat{B}_i = c_i^{+} - c_i$. Assuming $i<j$,
expanding the string
operator in terms of the $\hat{A}$ and $\hat{B}$ operators
and performing the adequate contractions, we get
$ <S_i^{+} S_j^{-} > =  D_{ij}/2$, with the determinant
\begin{equation}
D_{ij}=
\left|
\begin{array}{cccc}
G_{i,i+1} & G_{i,i+2} & ... & G_{i,j}\\
G_{i+1,i+1} & G_{i+1,i+2} & ... &G_{i+1,j}\\
... & ... & ... & ... \\
G_{j-1,i+1} & G_{j-1,i+2} & ... &G_{j-1,j}\\
\end{array}
\right|
,
\end{equation}
where
\begin{equation}
G_{i,j} = < \hat{B}_i \hat{A}_j > 
= \left( \sum_{\alpha \in FS} - \sum_{\alpha \not{\in} FS}\right)
\Psi^{\alpha}_i \Psi^{\alpha}_j
.
\end{equation}
This allows us to calculate numerically the $<<S_i^{+}
S_j^{-} >>$ correlations of the disordered XY chain.
The results are plotted on figure \ref{Fig5}, where
a cross-over to a $1/R^{2}$ behavior is visible.
Notice that in the $S^{+} S^{-}$ case,
the $1/R^{2}$ behavior is achieved for larger separations
than in the $S^{z} S^{z}$ case.
In addition, we have also calculated the equal-time single particle 
Green function, which also seems to decay as $1/R^2$ at large distances.

\section{Diagonalization of small clusters}
In order to have a further confirmation of the physical picture
that comes out from the previous analysis, we have
performed exact diagonalizations of small
clusters. 

\begin{figure}
\vspace{10cm}
\centerline{\epsfig{file=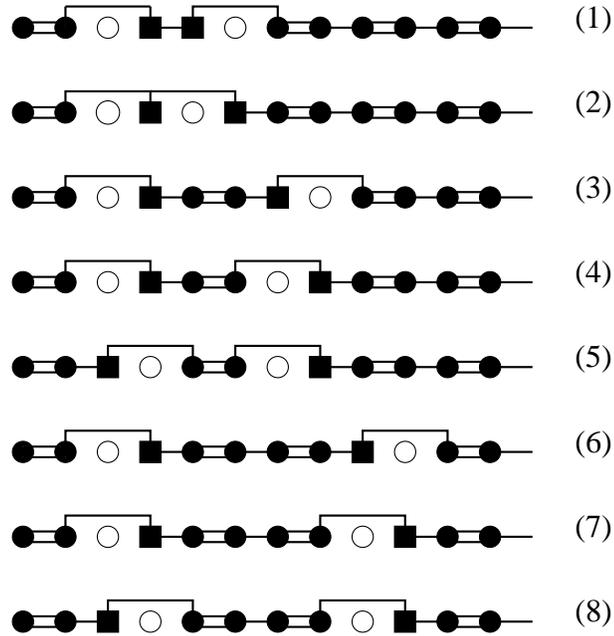,rwidth=10cm,rheight=10cm}}
\vspace{-5cm}
\caption{The eight inequivalent disorder realizations of the 
$12$ sites chain with $2$ impurities. The empty circles are
the impurities, the filled symbols are the spin $1/2$
sites, the filled squares being the solitons. A double
line stands for a strong bond ($J+\delta$) and
a single line stands for a weak bond ($J-\delta$).
}
\label{Fig6}
\end{figure}

We work on a chain of $12$ sites with $2$ non 
magnetic impurities. 
We enumerate the possible disorder realizations, which are shown on figure
\ref{Fig6}, and diagonalize
the spin Hamiltonian in all the $S_z$ sectors.
We also make the assumption that the domain wall created by each impurity 
in the dimerization pattern does not move, but remains trapped
at the impurity site. This simply implies that each impurity
releases a free spin which is located either at the right or at the left
of the impurity site (see the figure \ref{Fig6}).  
A strong bond has a value $J(1+\delta)$
and a weak bond $J(1-\delta)$. Since we work with a very small
chain, the dimerization should be large enough to make
the spin-Peierls correlation length much shorter
than the average spacing between impurities.
The Hamiltonian parameters
are thus far from being realistic. Moreover, we assume that the
second nearest neighbor interaction across the impurity
is equal to the weakest bond $J(1-\delta)$.
In practice, we take $J=1$ and $\delta=.6$.
The reason why we cannot address the problem for
realistic values of the parameters is that realistic
computations would involve a much larger number of spins, 
which is not accessible numerically. However, as we are going to show,
even our over-simplified system leads to some interesting results, 
not in disagreement with available experimental
data.

\begin{figure}
\vspace{10cm}
\centerline{\epsfig{file=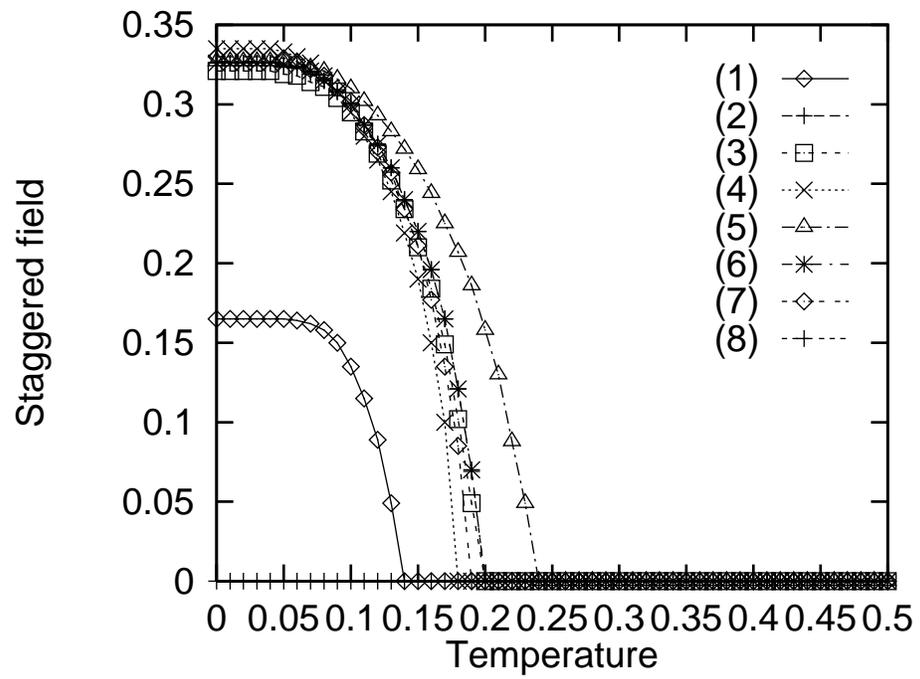,rwidth=10cm,rheight=10cm}}
\vspace{-5cm}
\caption{Staggered magnetic field as a funtion of temperature
for the eight disorder realizations
($J_{\perp}=0.315$).
}
\label{Fig7}
\end{figure}

In order to describe the antiferromagnetic transition we treat the
interchain coupling $J_\perp$ in a mean field approximation. 
Specifically, we impose a staggered magnetic field
$h_s$ and calculate the staggered magnetization
in the presence of $h_s$. We thus get the
values of the staggered magnetization $m_S(h_S,T)$
as a function of the staggered field and temperature $T$.
Next, we determine $h_s(T)$ self-consistently by imposing that
$h_S(T) = 2 J_{\perp} m_S(h_S,T)$. The self-consistent staggered
field is plotted on figure
\ref{Fig7} for the different disorder realisations. From that
figure we see that the on-set temperature at which a finite
staggered magnetization appears (in a sense the mean field
N\'eel temperature) is different for different 
disorder realisations. In particular, while most of the
realisations have almost the same on-set temperature, there is one
which has a larger temperature. This confirms our previous result
that magnetic correlations start to appear above the soliton
bandwidth and are due to rare disorder realisations which however dominate
in that temperature range. 

\begin{figure}
\vspace{10cm}
\centerline{\epsfig{file=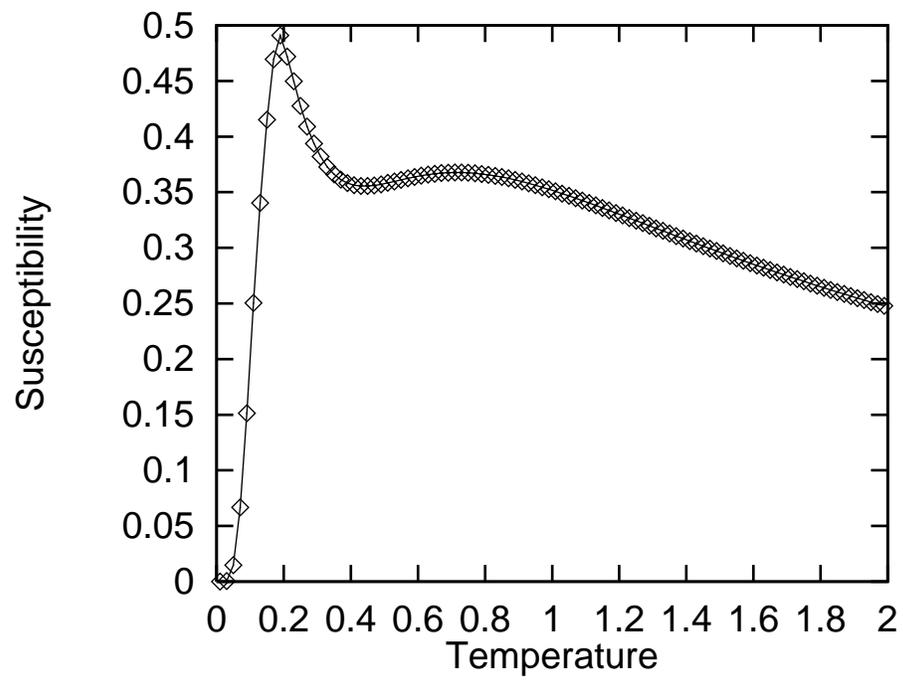,rwidth=10cm,rheight=10cm}}
\vspace{-5cm}
\caption{Susceptibility versus temperature ($J_{\perp}=0.315$).
}
\label{Fig8}
\end{figure}

Next, for each disorder realization, we have calculated the
susceptibility at the self-consistent point and
averaged over all the disordered
realizations. The average susceptibility as a function
of temperature is plotted on figure \ref{Fig8}. As the temperature
is decreased, the susceptibility starts to decrease for temperatures
smaller than the spin-Peierls gap. Instead of going monotonically
to zero for decreasing temperatures,
as in the case of a pure spin-Peierls system, the susceptibility
reaches a minimum, after which it start
increasing, and finally is cut-off by the antiferromagnetic
order. This behavior is in qualitative
agreement with the experimental data \cite{Si-a}.

\begin{figure}
\vspace{10cm}
\centerline{\epsfig{file=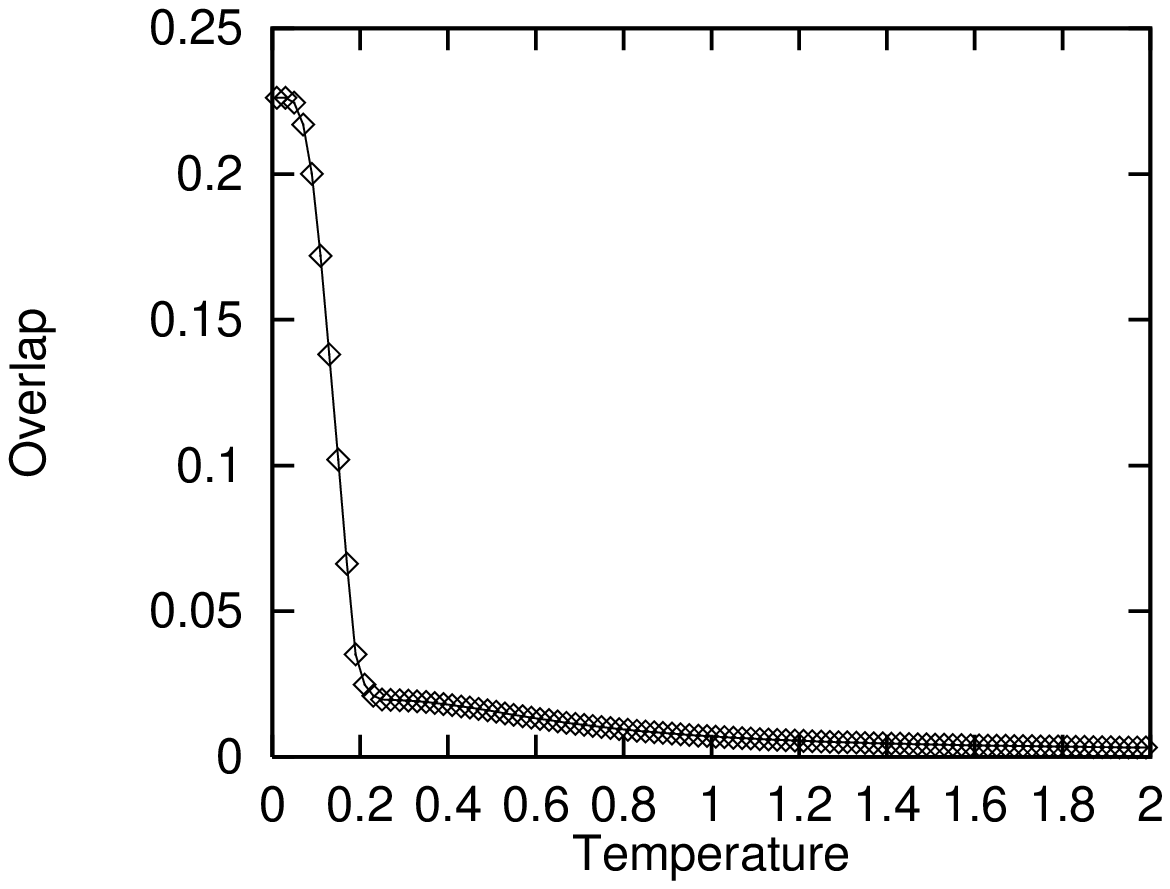,rwidth=10cm,rheight=10cm}}
\vspace{-5cm}
\caption{Variations of the overlap with the antiferromagnetic state
as a function of temperature ($J_{\perp}=0.315$).
}
\label{Fig9}
\end{figure}

We also plotted on figure \ref{Fig9} the variations of
the average overlap with the antiferromagnetic state.
If $|\Psi_{AF} \rangle$ denotes this state, then the overlap is
defined by
\begin{equation}
\frac{1}{\sum_{\alpha} \exp{(- \beta E_{\alpha})}}
\sum_{\alpha} |\langle \Psi_{AF} | \Psi_{\alpha} \rangle |^{2}
\exp{(- \beta E_{\alpha})}
,
\end{equation}
where $\alpha$ labels the eigenstates $|\Psi_{\alpha} \rangle$,
and $E_{\alpha}$ are the energies.
As expected, we observe a sharp increase of this overlap as the self
consistent staggered magnetic field is switched on. However, we also
observe an increase, although slow, at larger temperatures. 
This increase is consistent
with the existence of enhanced antiferromagnetic fluctuations
above the N\'eel temperature that were described in this paper.

\section{Discussion and conclusions}

In this article we have analyzed both analitycally and numerically 
a model for a doped spin-Peierls system. In this model a
major role is played by the localized spins which are released
by the impurities. These spins are antiferromagnetically coupled 
by the polarization of the singlet background. An important result that we 
find is that sensible 
antiferromagnetic fluctuations start to appear just below the spin-Peierls 
gap and well above the average 
value of the exchange coupling, which is related to
the coherence temperature of an ordered lattice made by those spins. 
In other words, there is a high temperature regime where the disordered
system is more correlated than an ordered one. 
In this temperature range, we find that the thermal correlation length
increases logarithmically with temperature. Moreover, in spite of the
presence of disorder, these fluctuations are quite long ranged and, in fact,
give rise at zero temperature to power low decaying spin-spin correlation 
functions, which has been numerically verified in the XY limit.

We believe that these disorder-induced magnetic fluctuations
are important in doped CuGeO$_3$, especially in driving the system
towards a N\'eel ordered phase. There are experimental evidences which,
in our opinion, reveal the importance of these fluctuations. 
For instance, the data of Ref.\onlinecite{Zn-b} show no saturation of 
the intensity of the antiferromagnetic Bragg peak as the temperature 
is lowered below $T_N$. This has been interpreted by the authors as a clear
signature of disorder. In addition, the extrapolation at 
$T=0$ of the value of staggered 
magnetization in the N\'eel state, which is 0.2$\mu_B$ for 4\% Zn-doping 
according to Refs.\onlinecite{Zn-b} and \onlinecite{Zn-c}, implies that not 
all the Cu spins are involved in the antiferromagnetism. 
Finally, we find interesting the uniform magnetization data $M(T)$ of 
Ref.\onlinecite{Si-a}, for Si doped compounds. In fact, for 
doping $\leq 0.005$, with decreasing temperature, $M(T)$ shows a first 
drop at $T_{SP}$, followed by a minimum, after which $M(T)$ starts 
increasing, without showing any saturation at low temperature. This is a 
clear evidence that doping induces some localized moments which are randomly 
distributed. For larger doping, the low temperature singularity is
cut-off by a second transition into a N\'eel phase. 
These results suggest that those localized moments do
participate actively
to the antiferromagnetic ordering. It is moreover suggestive in that 
Reference 
that the doping concentration at which $T_N$ is maximum seems to coincide 
with the vanishing of $T_{SP}$, or better the closing of the 
spin-Peierls gap, as if, in the presence of an underlying spin-gap,
doping helps antiferromagnetism, which is instead reduced 
by doping in the absence of the gap. This result is compatible with
our model.  

Unfortunately, we can not give a realistic estimate of the N\'eel 
temperature within our single-chain toy model. This would need the 
inclusion of interchain
couplings which can not be handled by the RG approach we have used. 

It is now worthwhile to discuss more in detail our theoretical motivations
in support of these disorder-induced magnetic fluctuations.
As we said in the introduction, an alternative explanation of the
appearence, upon such a weak doping, of a magnetically ordered phase, is that
already the pure compound is quite close to the quantum critical
point separating the spin-liquid phase, with a spin-gap in the
excitation spectrum, from the N\'eel phase, with gapless spin wave
excitations. Then one can imagine that doping effectively 
decreases the dimerization or increases the interchain coupling so to push
the system in the N\'eel phase. However, this simple view
is in our opinion not fully satisfactory, as we are going to discuss.

Katoh and Imada\cite{Imada} have recently analyzed by quantum 
Monte Carlo
and spin wave theory a quasi one-dimensional Heisenberg model with
a dimerized exchange. Their model differs from the one expected to
describe the CuGeO$_3$ for the absence of an intra-chain next-nearest
neighbour coupling, as well as for the fact that in this compound
the dimerization is staggered in the transverse direction, contrary
to the model analyzed by those authors. Anyway, what Katoh and Imada
find is that, for a realistic value of the ratio of the interchain to 
intrachain
exchange ($\sim 0.1$), the critical value of the dimerization
separating the N\'eel from the spin-liquid phase is $\delta_c\simeq 0.05$,
not far from the estimated value of $\delta=0.03$ in the pure CuGeO$_3$.
What is more important for what we are going to say, is that spin-wave
theory overestimates $\delta_c$, giving a value $\simeq 0.7$.
If one does spin-wave theory for a realistic model, taking into
account that the dimerization is staggered in the transverse direction
and including a next-nearest neighbour coupling, one finds that
a model describing CuGeO$_3$ is well inside the spin-liquid phase,
and it would need a really large modification of the Hamiltonian parameters
to enter the N\'eel phase. One could imagine that
the quantum fluctuations not present in the
simple spin-wave theory might change these results and push the
system towards the N\'eel phase. However these quantum fluctuation 
corrections go in the opposite direction in the case
analyzed by Katoh and Imada, and we do not understand why they should
work oppositely in the realistic case. 

There have been other attempts, which make use of bosonization, to
derive the phase diagram of a quasi-one dimensional Heisenberg
model with dimerization and intra-chain next-nearest neighbour
exchange\cite{Fukuyama,Bishop}. In Ref.\onlinecite{Bishop}
it is claimed that, within this approach it is possible to show that
the model describing the pure CuGeO$_3$ is indeed very close to
the transition towards a N\'eel phase. However, also these approaches
are not fully satisfactory in our opinion. First of all, as a
consequence of the one-dimensional treatment (the interchain coupling is
treated in mean-field, assuming that the neighbouring chains provide
a staggered magnetic field, which is then calculated self-consistently),
the transition between the spin-liquid phase and the N\'eel phase
is accompanied by the vanishing of the dimerization. This is not what
is seen experimentally (see e.g. Ref.\onlinecite{Zn-c}). In fact,
as we already said, a N\'eel phase is not incompatible with a
staggered exchange in more than one dimension.
A simple spin-wave calculation shows that there is a gain in energy
by switching on a small dimerization $\delta$, given by
\begin{eqnarray*}
E(\delta)-E(0) &=& \frac{1}{2V} \sum_{\vec{k}}
\sqrt{ (J+J_y-2J'\sin^2 k_x)^2 - (J\cos k_x + J_y \cos k_y)^2
- J^2\delta^2\sin^2 k_x}\\ 
&-& 
\sqrt{ (J+J_y-2J'\sin^2 k_x)^2 - (J\cos k_x + J_y \cos k_y)^2},
\end{eqnarray*}
where $J(1\pm \delta)$ is the intra-chain nearest neighbour 
exchange for bonds of different parity, $J_y$ is the interchain
exchange (we have assumed a two-dimensional system, since the
coupling in the other direction is much smaller than
$J$ and $J_y$), and $J'$ in the intra-chain  next nearest neighbour
exchange.
This gain is quadratic in $\delta$, for small $\delta$, and 
should be compared with the loss of elastic energy due to the 
lattice distortion, which is also quadratic. Therefore,
it is not impossible that a N\'eel state is stable in spite of 
a dimerized exchange. In addition, the
bosonization approach is extremely delicate and not fully justifiable 
for a $J'\sim 0.23 J$,
very close to the value $0.2411$ which separates the 
gapless Luttinger liquid phase from the gapped valence bond regime.

In conclusion, we believe that one really needs to take into account
those disorder induced magnetic fluctuations to explain why the
spin-liquid phase of the spin-Peierls compound CuGeO$_3$ is so
unstable upon doping, and easily gives way to a N\'eel phase.
The analysis presented here is a first attempt in this
direction. 

\section{Acknowledgements}
We thank J. Lorenzana for useful discussions. This 
work has been partly supported by INFM, project HTSC.

\begin{references}
\bibitem{nota-imp} The  
correlation between the position of the impurities and
of the solitons has been revealed in neutron scattering 
experiments on weakly doped samples in the presence of a strong magnetic 
field able to excite solitons\cite{soliton}.
\end{references}
\end{document}